# Production of $^{82}$Se enriched Zinc Selenide (ZnSe) crystals for the study of neutrinoless double beta decay

Authors:   I. Dafinei (1), S. Nagorny (3), S. Pirro (2), L. Cardani (1), M. Clemenza (5 and 6),
F. Ferroni (1 and 4), M. Laubenstein (2), S. Nisi (2), L. Pattavina (2), K. Schaeffner (3),
M. L. di Vacri (2), A. Boyarintsev (7), I. Breslavskii (7), S.Galkin (7), A. Lalayants (7),
I. Rybalka (7), V. Zvereva (7), M. Enculescu (8)

[1]INFN - Sezione di Roma, Roma I-00185 - Italy
[2]INFN - Laboratori Nazionali del Gran Sasso, Assergi (L'Aquila) I-67010 – Italy
[3]INFN - Gran Sasso Science Institute, 67100, L'Aquila - Italy
[4]Dipartimento di Fisica, Sapienza Università di Roma, Roma I-00185 – Italy
[5]Dipartimento di Fisica, Università di Milano-Bicocca, Milano I-20126 - Italy
[6]INFN - Sezione di Milano Bicocca, Milano I-20126 - Italy
[7]Institute for Scintillation Materials, National Academy of Sciences of Ukraine, Lenin ave. 60, Kharkiv, 61001, Ukraine
[8]National Institute of Materials Physics, Atomiștilor Str. 405A, 077125, Măgurele, România

[a]Corresponding author: I. Dafinei; INFN, Sezione di Roma; P-le Aldo Moro 2, 00185-Roma, ITALY
Tel.: +390649914333; fax: +39064454835; e-mail: ioan.dafinei@roma1.infn.it

*Abstract*
*High purity Zinc Selenide (ZnSe) crystals are produced starting from elemental Zn and Se to be used for the search of the neutrinoless double beta decay (0vDBD) of $^{82}$Se. In order to increase the number of emitting nuclides, enriched $^{82}$Se is used. Dedicated production lines for the synthesis and conditioning of the Zn$^{82}$Se powder in order to make it suitable for crystal growth were assembled compliant with radio-purity constraints specific to rare event physics experiments. Besides routine check of impurities concentration, high sensitivity measurements are made for radio-isotope concentrations in raw materials, reactants, consumables, ancillaries and intermediary products used for ZnSe crystals production. Indications are given on the crystals perfection and how it is achieved. Since very expensive isotopically enriched material ($^{82}$Se) is used, a special attention is given for acquiring the maximum yield in the mass balance of all production stages. Production and certification protocols are presented and resulting ready-to-use Zn$^{82}$Se crystals are described.*
______________________________________________________________________
______________________________________________________________________

## 1. Introduction

Many of the unsolved problems in fundamental physics are related to rare event processes, so called either because they effectively have a low probability to happen (double-beta decay and other rare nuclear decay modes) or because the particles involved have an extremely low interaction cross-section with matter (neutrinos or hypothetical particles like axions, WIMPs, etc). Whatever the nature of such processes, the common conditions for their experimental proof are: (i) develop the experimental conditions which guarantee an extremely low (ideally zero) background; (ii) build extremely high sensitivity detectors. Related to the first constraint, the experimental facilities have to be sited deep underground or under sea (to reduce the impact of cosmic radiation) and the detector material and immediate surroundings have to be particularly radio-pure (to reduce the level of natural radioactivity). For the construction of new, high sensitivity detectors, very often crystals are used as e.g. for the cryogenic bolometer technique. The crystals used in this case, besides the low concentration of radio-active impurities (typically below $10^{-13}$ g/g), have to satisfy the need of a high crystal perfection in order to allow for a high sensitivity detector.

The neutrinoless double beta decay (0vDBD) is a very rare process [1, 2] of particular importance since if verified experimentally, it will be the direct proof that neutrino is a massive Majorana particle and that the lepton number conservation can be violated in natural processes. Besides that, the measurement of the half-life of this process would give a hint on the neutrino absolute mass scale and on the neutrino mass hierarchy. The cryogenic bolometer technique is one of the promising experimental approaches in 0vDBD experiments because of the achievable excellent energy resolution and extremely high detection efficiency [3]. A cryogenic bolometer consists of an absorbing material cooled to very low temperature (<100mK) and a thermometer able to measure the temperature rise due to the tiny energy deposit caused by a particle interaction in the absorber. The drawback of the method is that a priori it cannot discriminate between different particles which release the same amount of energy in the detector. Scintillating bolometers are devices in which the absorbing material is a scintillator and the light emitted by the energy deposit of the particle is also measured by a separated light detector [4]. Simultaneous measurement of bolometric and scintillation signals allows for the discrimination between alpha and beta particles thus drastically reducing the background [5] (virtually to a zero level as the amount of detected scintillation light depends on the particle type). The production of crystals to be used as scintillating bolometers becomes particularly burdensome when enriched (expensive) materials are used, because of the additional demand of a high throughput yield in all stages of the manufacturing process.

The current article reports on the production of high purity $Zn^{82}Se$ crystals for the study of the neutrinoless double beta decay of $^{82}Se$ using scintillating bolometers. A detailed description is given of all production phases, from the synthesis of $Zn^{82}Se$ powder to the growth of $Zn^{82}Se$ crystals and their mechanical processing to obtain of ready-to-use scintillating bolometers. Special attention was paid to ensure a maximum yield in the mass balance for all stages of the crystal production process.

## 2. $Zn^{82}Se$ crystals for cryogenic experiments

Several studies were dedicated to the choice of the best suited nuclide and crystal host to be used as scintillating bolometer in a 0vDBD experiment [6] and the use of enriched isotopes for the production of such crystals was widely discussed [7]. In principle, any crystal containing a 0vDBD isotope can be used in a cryogenic experiment for evidencing this very rare decay. This is e.g. true for ZnSe since $^{82}Se$ is one of the 0vDBD candidates. ZnSe crystal has a privileged position due to good thermal and mechanical properties and also due to the relatively high value of the decay energy of $^{82}Se$ (2.998 MeV), above the most intense natural gamma line (2.615 MeV) arising from the $^{208}Tl$ decay. The relatively low natural abundance of $^{82}Se$ (8.73%) can be compensated by enrichment and last but not least, ZnSe crystal is an intrinsic scintillator thus can be used as scintillating bolometer. The LUCIFER project [8] was the first attempt to reach a background-free condition in a 0vDBD experiment by using scintillating crystals grown from isotopically enriched materials and $Zn^{82}Se$ crystal was the first candidate for this experiment [9]. The guidelines for the production of crystals to be used in a 0vDBD search were drawn in a previous work dedicated to $TeO_2$ crystals used in CUORE experiment [10]. However, in that case the industrial production was already in place and the work consisted in implementing those modifications aimed at ensuring the harsh specifications imposed by CUORE (higher crystal perfection and radio-purity). The production of scintillating crystals is a more complex case since no scintillating material suited for 0vDBD experiments was produced in large scale till today. It was necessary therefore to reconsider the entire production process of $Zn^{82}Se$ crystals in accordance with the specific requirements of low-background 0vDBD experiments.

### 2.1. Zinc Selenide (ZnSe) crystal properties

Zinc selenide is rarely found in nature and the synthesis of ZnSe compound can be made in different forms (powder, bulk pieces or thin layers) by different methods and starting from different raw materials, depending on the use that is given to the final product. ZnSe can be produced also in form of crystals which are grown by different methods and can be made in both hexagonal (wurtzite) and cubic (zincblende) crystal structure. Zinc selenide is a II-VI semiconductor with a relatively large bandgap (about 2.70 eV at 25 °C, value strongly dependent on the production conditions) which is used as an infrared optical material because of its wide range of transmission (0.45 μm to 21.5 μm). The material is also used as lasing or LED medium and as scintillator, especially in X-ray and gamma ray detectors in which case can also be doped for increased scintillating performance. Table 1 gives the general properties of ZnSe crystal.

Table 1 General properties of ZnSe crystal

| Crystal type | cubic |
|---|---|
| Lattice constant | a=5.657Å |
| Density | 5.27 g/cm$^3$ at 25 °C |
| Melting point | 1525 °C |
| Refractive index (n) | 2.417 - 2.385 at 8 - 13 µm<br>2.40272 at 10.6 µm |
| dn/dT | +61 x 10$^{-6}$/°C at 10.6 µm at 298K |
| Transmission range | 0.45 to 21.00 µm |
| Band gap (at 10K) | 2.82 eV (direct) |
| Bulk absorption coefficient | 5·10$^{-4}$ cm$^{-1}$ at 10.6 µm |
| Reflection loss | 29.1% at 10.6 µm (2 surfaces) |
| Young's Modulus | 6.72·10$^9$ dynes/mm$^2$ |
| Specific Heat at 25 °C | 356 J/kg/°C |
| Linear thermal expansion | 7.57·10$^{-6}$ / °C at 20 °C |
| Hardness (Knoop) | 105-120 kg/mm$^2$ |
| Solubility | 0.001g/100g water |
| Molecular Weight | 144.37 (8.73% $^{82}$Se, natural)<br>147.25 (95% $^{82}$Se, enriched) |

Although it has several applications in IR optics and optoelectronic devices, ZnSe is quite seldom produced in the form of large (single) crystals. For an effective 0vDBD experiment based on the cryogenic bolometer technique, crystals of relatively large dimensions (> 400 g) are needed. The chemical vapor deposition (CVD) technique (the most efficient method for obtaining large samples at a convenient price) was discarded in early phases of our project due to the poor scintillation and bolometric performance of ZnSe samples grown by this method. Such samples, in spite of their good optical aspect (transparent, homogeneous, no visible bubbles or inclusions) generally have a polycrystalline structure which causes the deterioration of the bolometric performance and the quenching of scintillation at low temperatures. The adverse effect of polycrystalline structure and extended defects was observed also in crystals grown from melt by different versions of the Bridgman technique.

The growth of large ZnSe crystals from the melt is very difficult due to the high melting point (1525°C) and total vapor pressure (0.2MPa at 1525°C) and is further hampered by the high volatility of the components Zn and Se, which in addition have different vapor pressures leading to a deviation from the stoichiometric composition of the melt during the growth process. The very narrow homogeneity range in the ZnSe phase diagram [11, 12] and the solid-solid phase transition from hexagonal wurtzite to cubic sphalerite (zinc blende) occurring in the cooling phase of the grown crystal at 1425 °C [13, 14] are additional problems to be solved when the melt growth option is considered to obtain large, homogeneous ZnSe crystals. All crystals discussed in this work were grown from the melt in graphite crucibles using a modified Bridgman technique.

### 2.2. Specific requirements for rare events physics application

The crystal absorber forms the core of a cryogenic bolometer. The energy deposited by a particle is measured through the consequent temperature increase of the crystal. In 0vDBD experiments the crystal is both the source of the decay and the bulk of the detector. A high level of crystal perfection and homogeneity are requested for high sensitivity bolometers because the presence of bulk defects like bubbles and metallic inclusions or any other dishomogeneity like veils and cracks may drastically reduce the signal amplitude and the energy resolution of such detectors. There are several studies dedicated to the perfection of ZnSe crystals [15-17]. The crystals grown in this work were subject to a special thermal regime aimed at reducing the presence of such defects. The crystal cooling process was conducted under conditions that did not alter the crystal perfection during the phase transition from wurtzite to sphalerite symmetry.

The capability of the cryogenic technique to identify the DBD signal is limited by the presence of environmental radioactivity and radioactive contaminants in the detector. Therefore the detector has to be free of any contaminant that can mimic or deface the DBD signal by producing an energy deposit in the region of interest for the 0vDBD. The contamination may come from long-lived, naturally occurring isotopes, such as $^{238}$U, $^{232}$Th, $^{40}$K and their daughters and from cosmogenic activation of the detector materials and surroundings. In the case of Zn$^{82}$Se crystals, this means a detailed monitoring of all materials, tools and facilities used for the synthesis of Zn$^{82}$Se powder and the crystal growth, the chemical and mechanical processing of grown crystals, their handling, packing and storage before mounting in the experimental setup. Land transport and underground storage of raw materials and finished crystals are necessary in order to minimize cosmogenic activation [18].

Another important feature required for scintillating bolometers is a good optical transmittance. Highly transparent crystals are needed in order to avoid reabsorption of scintillation light. Zinc selenide is a direct-gap semiconductor in which the normal band-to-band optical absorption is strongly perturbed by absorption on excitons and impurities. The concentration of impurities has to be reduced (typically below ppm levels) also because they may alter the transmission in the IR region of the spectrum usually dominated by the absorption on free carriers. A reduced IR transmission may alter the propagation of phonons with consequent reduction of bolometric performance of the crystal.

### 3. Experimental details

For the Zn$^{82}$Se crystals in this work, dedicated equipment was developed for all phases of the production process mainly because of the very strict radio-purity requirements. Details will be given in the corresponding paragraphs. The instrumentation used for the certification of Zn$^{82}$Se crystal and intermediate products was:

   a) **ICPMS (chemical purity and radio-purity analysis, isotopic concentration measurements)**

Inductively Coupled Plasma Mass Spectrometry (ICP-MS) measurements were performed at Istituto Nazionale di Fisica Nucleare (INFN), Laboratori Nazionali del Gran Sasso (LNGS) on all raw materials and intermediate products. A quadrupole mass spectrometer from Agilent Technologies, model 7500a (2001), was used with dedicated tuning of the machine in order to have high sensitivity, stable signals and reliable results concerning matrix effects, especially for the radio-purity measurements. A Babington-type nebulizer was used, which supports a high concentration of dissolved solid without any clogging. Dedicated sample preparation protocols (different procedures depending on the nature of samples) were developed to reach the best possible sensitivity while limiting the salt concentration in the solution to be measured. The sensitivities reached were $2 \cdot 10^{-10}$ g/g for $^{232}$Th and $^{238}$U in solid raw materials and consumables, $10^{-12}$ g/g in H$_2$O and $10^{-11}$ g/g in all other reagents.

   b) **HPGe gamma-ray spectrometry (radio-purity analysis)**

High-resolution gamma-ray spectrometry with cooled hyperpure germanium (HPGe) crystal detectors was used for the precise measurement of radioactive contamination of elemental Zn and Se. HPGe spectroscopy measurements reported in this work were performed at the low-radioactive laboratory of the LNGS underground site [19]. All detectors are shielded from the environmental background with lead shields, specially selected for low levels radioactive contamination, and surrounded with special boxes, to minimize radon contamination. The measurement setup may reach sensitivities of $10^{-12}$ g/g for the U and Th chains when large samples are measured for a very long acquisition time. Data reported in this article are calculated at 90% CL.

   c) **Electron microscopy (crystal defects analysis)**

The morphology of ZnSe crystal samples was evaluated using an EVO 50 XVP Carl Zeiss microscope with LaB$_6$ filament, 2.0 nm resolution at maximum accelerating voltage of 30kV. The microscope is equipped with both secondary electrons detector (SE) and backscattered electrons detector (BSD). Elemental composition of samples can be analyzed with EDX (Energy-dispersive X-ray analysis) accessory, Quantax Bruker 200 model, Peltier cooled X-ray detector, energy resolution of 0.133 keV for Mn K$\alpha$ lines (5.898 keV and 5.887 keV).

### d) Optical transmission

A commercial spectrophotometer model Shimadzu UV mini-1240 operating in the range 200 - 1100 nm with a 1 nm resolution was used for transmission spectra in the visible region and Shimadzu Fourier Transform Infrared spectrophotometer IR Affinity-1 operating in the range 7800-350 cm$^{-1}$ (1.3 - 28.5 µm) with a 2 cm$^{-1}$ resolution for the IR region were used at ISMA Kharkiv, Ukraine. The samples were optically polished on plan parallel faces using the same technique for all samples thus allowing for a comparative study of optical transmission in the range 480 nm – 25 µm. Very important information is received from the measurement of the absorption intensity in the regions 550-650 nm (point defect absorption) and 3-15 µm ($Fe^{2+}$, CSe and $CSe_2$ absorption).

### e) Cryogenic test

The $^3$He/$^4$He dilution refrigerator of hall **C**, at LNGS [20], was used for cryogenic tests aimed at checking the bolometric and scintillation performance of $Zn^{82}Se$ crystals in terms of energy resolution, background rejection capability and intrinsic radio-purity (µBq/kg in $^{232}$Th, $^{238}$U and their daughters).

## 4. $Zn^{82}Se$ crystal production

The different steps of $Zn^{82}Se$ crystal production starting from elemental zinc and selenium are illustrated in figure 1. The elemental ultrapure Zn was produced by the National Science Center "Kharkov Institute of Physics & Technology" (KIPT) in Kharkov Ukraine while the enriched Se was delivered by the URENCO Stable Isotopes unit based at Almelo in the Netherlands. The synthesis of ZnSe powder and its conditioning for crystal growth (steps 11, 12 and 14) were made at ISMA Kharkiv, Ukraine and the processing of the crystal ingots at LNGS, Italy. The certification process was developed and performed in INFN laboratories. The measurement methods for radio-purity certification as well as the protocols applied for each production phase including handling, cutting and polishing of $Zn^{82}Se$ crystals in this work are similar to those described in [9] for the $TeO_2$ crystals produced for the CUORE experiment. Given the very high price of the material used, in particular for the enriched selenium, additional measures have been taken to minimize the loss of material in all stages of production. Recovery of the material loss at cutting and polishing was also foreseen (step 17).

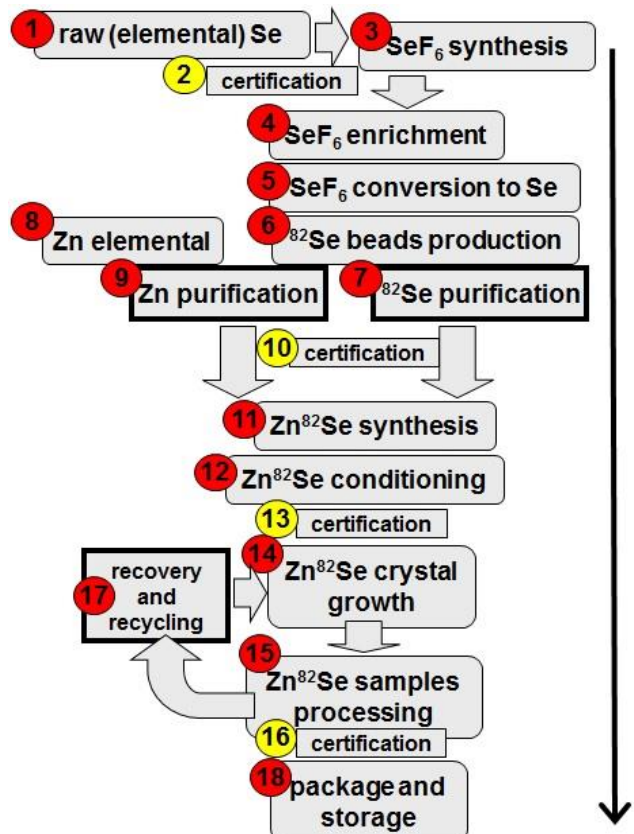

Fig. 1 Diagram of $Zn^{82}Se$ crystal production in this work

### 4.1. Production of elemental Zn and Se

The elemental ultrapure Zn was produced at KIPT Kharkov in Ukraine [21] starting from Zn metal which is commercially available and by applying a dedicated purification process using multi-stage vacuum distillation. The final Zn metal product was delivered in form of beads.

**Table 2** Chemical purity expressed in ppb as measured by ICP-MS on samples of enriched $^{82}$Se and Ze randomly chosen from regular production at URENCO and at KIPT respectively

| element | $^{82}$Se initial | $^{82}$Se distilled | Zn initial | Zn purified |
|---|---|---|---|---|
| Na | 500000 | <1000 | 1000 | 30 |
| Mg | <1000 | <100 | | |
| Al | | <200 | 4000 | 35 |
| Si | | 18000 | 7000 | 40 |
| S | 1140000 | 130000 | | |
| K | <2500 | <2000 | 3000 | 35 |
| Ca | | 5000 | 2000 | 70 |
| V | <20 | <40 | <200 | <9 |
| Cr | 100 | <100 | <200 | <90 |
| Mn | <100 | <10 | <200 | <90 |
| Fe | 700 | 110 | 30000 | 56 |
| Co | <10 | <4 | <200 | <10 |
| Ni | 400 | 64 | <300 | <90 |
| Cu | 250 | <10 | 10000 | 110 |
| Mo | <300 | <10 | <2000 | 4 |
| Cd | | <4 | 20000 | 4300 |
| Sn | | | 15000 | <300 |
| Tl | | | <2000 | 180 |
| Pb | | <6 | <3000 | 290 |
| Bi | | 6 | <1000 | <2 |
| Th | <0.4 | <0.1 | <2000 | <0.2 |
| U | <0.2 | <0.1 | <2000 | <0.2 |

The production of $^{82}$Se-enriched selenium starting from elemental selenium turned into $SeF_6$ for enrichment by gas centrifugation and further converted into selenium was described elsewhere [22, 23]. Supplementary purification of the enriched Se (step 7 in fig.1) was applied at the URENCO production site using a distillation procedure which cannot be disclosed. The chemical purity of the enriched $^{82}$Se from URENCO are given in table 2 together with the purity of Zn produced at KIPT. Besides the reduction of sodium and sulfur concentration which was the main purpose of the supplementary purification of $^{82}$Se, the reduced concentration of Cr, Fe, Ni and Cu is to be noted. In the case of Zn, a decrease of almost two orders of magnitude is noticed for all the concentrations of contaminants. The isotopic concentration of the enriched Se beads delivered by URENCO and used in the current work are given in table 3.

**Table 3** The isotopic concentration of the enriched Se beads used in the current work

| $^{74}$Se (%) | $^{76}$Se (%) | $^{77}$Se (%) | $^{78}$Se (%) | $^{80}$Se (%) | $^{82}$Se (%) |
|---|---|---|---|---|---|
| <0.01 | <0.01 | <0.01 | <0.01 | 3.67 | 96.3 |

The radio-purity of Zn and enriched Se (before and after the distillation) was measured by HPGe gamma spectroscopy at LNGS and the results are given in table 4. The extremely low-background conditions in which the measurement was carried out and the high radio-purity of the $^{82}$Se allowed us to establish for the double-beta decay of $^{82}$Se to excited states of $^{82}$Kr the most stringent lower limits on the half-lives of the order of $10^{22}$ y, with a 90% C.L. [24].

**Table 4** Radiopurity of enriched $^{82}$Se and natural Zn. Concentration of radioactive impurities are given in mBq/kg.

| contaminant | Se (enr.) | Se (enr.&pur.) | Zn (nat.) |
|---|---|---|---|
| $^{238}$U / $^{226}$Ra | <0.41 | <0.11 | <0.07 |
| $^{238}$U / $^{234}$Th | <27.00 | <6.20 | <6.20 |
| $^{232}$Th / $^{228}$Th | 1.4 ± 0.2 | <0.11 | <0.04 |
| $^{232}$Th / $^{228}$Ra | <0.37 | <0.06 | <0.09 |
| $^{40}$K | 3 ± 1 | <0.99 | <0.38 |
| $^{60}$Co | <0.17 | <0.07 | <0.04 |
| $^{235}$U | <0.30 | <0.07 | <0.09 |
| $^{137}$Cs | <0.08 | <0.01 | <0.03 |

### 4.2. Synthesis of Zn$^{82}$Se powder and conditioning for crystal growth

The synthesis of Zn$^{82}$Se powder and its conditioning for crystal growth were made using ultraclean fused quartz reactors (Heraeus products HSQ900 and HSQ910). The whole process consisted of three distinct steps which are described in the following.

In the first step (primary synthesis: S-1), the synthesis of Zn$^{82}$Se is made in vapor phase by evaporation of Zn and enriched $^{82}$Se in Ar atmosphere at 950°C. The analysis of the ZnSe powder obtained in this first step (S-1), showed that not all the material reacted. Roughly 222 g of enriched $^{82}$Se and roughly 178 g of Zn are processed in a large ampule to make the first step. Approximately 90% of the crop is selected as first quality material (containing 95% Zn$^{82}$Se and 5% not fully reacted material). The remaining 10% (containing approximately 70% Zn$^{82}$Se and 30% material not fully reacted) is collected separately as second quality material. The material resulted from several S-1 runs is re-processed in a so-called Vertical Thermal Treatment (VTT) aimed at completing the synthesis of material not fully reacted in S-1. In the VTT procedure the powder collected in several S-1 crops (first and second quality) is conveniently put together in a dedicated ampoule and thermally treated at 950 °C in Ar atmosphere. The yield of the VTT process is of the order of 99.60% first quality material. The remaining 0.40% of unreacted (waste) material is recovered and thermally treated in the following VTT run. The material (first quality) resulted after the VTT proved to be still not suitable for efficient crystal growth because it still contains parts of not fully reacted Zn and $^{82}$Se. During the crystal growth process, possible elemental Zn will evaporate and escape the crucible. Elemental selenium instead if present, will react with the graphite crucible and give Carbon Diselenide (CSe$_2$) which is then imbedded in the crystal as impurity. In order to avoid this problem and guarantee a good quality powder for the crystal growth, the powder resulted after VTT was subject to a third treatment, so called Horizontal Thermal Treatment (HTT). The HTT is performed in a dedicated reactor with two different chambers allowing for a gas flow to flush away (or let them further react) free Zn and Se. The scheme of the reactor used for HTT is given in fig. 2. The presence of two different reactor volumes prevents the contamination of the purified material by the unreacted material flushed away by the gas flow. During a HTT run roughly 530 g of Zn$^{82}$Se resulting from the VTT process is treated. The treatment consists of slowly heating the powder under hydrogen flow up to 650°C. During the heating, starting at 300 °C, H$_2$$^{82}$Se will form and will be partially purged (in gaseous form) out of the small ampoule. Note that the H$_2$ flow is rather weak avoiding the extraction of all the free selenium. After a few tens of minutes, the temperature is increased enough and reaches 500 °C when a new reaction will take place:

$$H_2{}^{82}Se + Zn \rightarrow Zn^{82}Se + H_2 \qquad (1)$$

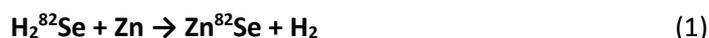

This regime is maintained at 650 °C (the temperature that maximize the reaction) in order to ensure an exhaustive synthesis of Zn and $^{82}$Se. In fact, other mechanisms enter in the reaction. Selenium melts at 221 °C and boils at 685 °C. This means that during the reaction part of the selenium will evaporate and escape the ampoule. Once all the free selenium is completely escaped (or reacted) we have to eliminate the residual zinc. Excess zinc will be present for two reasons:

a) Zn has a higher melting and boiling point with respect to Se, which means that in all previous treatments (S-1 and VTT), the loss of selenium is larger with respect to that of zinc
b) During HTT process $^{82}$Se will be purged from the ampoule in form of $H_2^{82}$Se or due to evaporation

In order to remove the residual Zn from the Zn$^{82}$Se powder, the temperature is slowly raised just above the boiling point of Zn. During this procedure, argon is purged in the system instead of Hydrogen. The atomic mass of Ar is much closer to the one of Zn which makes it more efficient as scavenger for the Zn to be taken out from the ampoule. Moreover, argon is preferred for being much less dangerous than $H_2$. The control of different gas flux in the system is made through a network of pipes and valves illustrated in fig. 2. The inner ampoule (containing the Zn$^{82}$Se powder) is continuously rotated by a step motor. A dry vacuum pump (Edwards nXDS6i 100) is used to pump the whole reactor before starting the thermal treatment. The transfer of the Zn$^{82}$Se powder inside and outside the reactor is made in argon atmosphere through a system which avoids the contact of the powder to free atmosphere. The yield of the HTT process is of the order of 99.40% first quality material, ready for crystal growth. The total yield of the process of Zn$^{82}$Se powder synthesis and conditioning for crystal growth was 98.35% (99.55% at S-1, 99.40% at VTT and 99.40% at HTT).

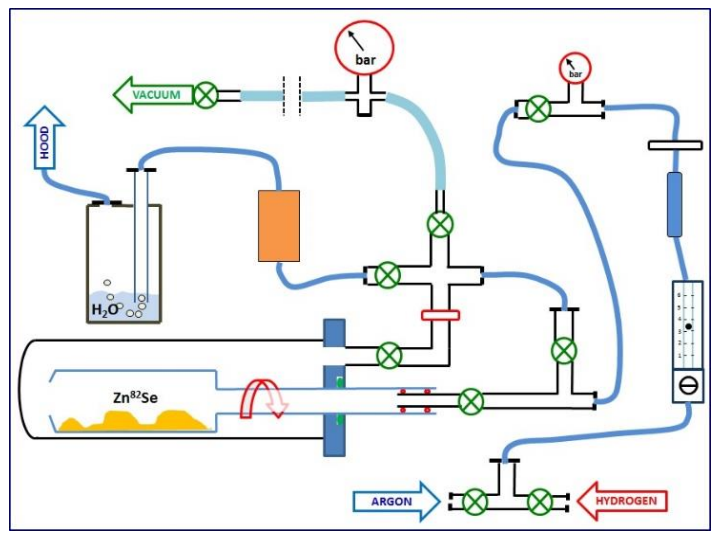

Fig. 2 Schematic view of the reactor used for the Horizontal Thermal Treatment (HTT)

### 4.3. Zn$^{82}$Se crystal growth

All crystals discussed in this work were grown from the melt, in graphite crucibles using the Bridgman technique. Loading crucibles with Zn$^{82}$Se powder was performed in controlled atmosphere (ultrapure Ar, relative humidity <1%) using a glove box and other dedicated equipment, previously certified for radio-purity (gloves, containers, plastic bags, jars and pestles, etc). The crystal growth was performed in a vertical furnace kept in argon atmosphere (up to 15MPa). The furnace [25] was composed of two heaters giving a temperature gradient designed to go from 1575°C to 900°C in the melt zone and further to insure a temperature of about 900°C in the thermalization zone. The growth rate was typically 1 mm/hour and the accuracy of temperature control ±3°C. The whole process (temperature control and crucible movement) was computer controlled using a software of ISMA Kharkiv property. The thermalization in the isothermal region of the heater at about 900° C, followed by gradual cooling to room temperature was applied to reduce the residual thermal stresses in the crystal. Zn$^{82}$Se crystal growth process was prone to big difficulties due to technical problems arising from the need of growing large ZnSe crystals with a low coefficient of material loss. Losses during crystal growth is due to thermal dissociation and due to diffusion through the walls of the graphite crucible. Further complication comes from the fact that evaporation and diffusion are different for the two elements, Zn and Se: (i) evaporation of the decomposed melt (Zn and Se$_2$) is different due to the fact that partial pressure of Zn is twice as much as the partial pressure of Se$_2$; (ii) the diffusion through the walls of graphite crucible is different due to the larger dimension of the Se$_2$ molecule with respect to Zn atom. In order to minimize Zn$^{82}$Se loss the graphite crucibles were longer than usual (about 250 mm including the seed compartment) and sealed with a screwed cap. The evaporated material deposited inside the crucible on the upper side of the walls (approximately 12% of initial charge) as well as the material evaporated and condensed in powder form on the cold parts of the furnace (about 5% of initial charge) was carefully collected for future recycling. The as-grown ingot represented typically 78% of the initial charge, i.e. the total yield in terms of mass balance was approximately 95%. Fig. 3 gives a typical $Zn^{82}Se$ crystal ingot ready for mechanical processing (cutting, shaping and polishing).

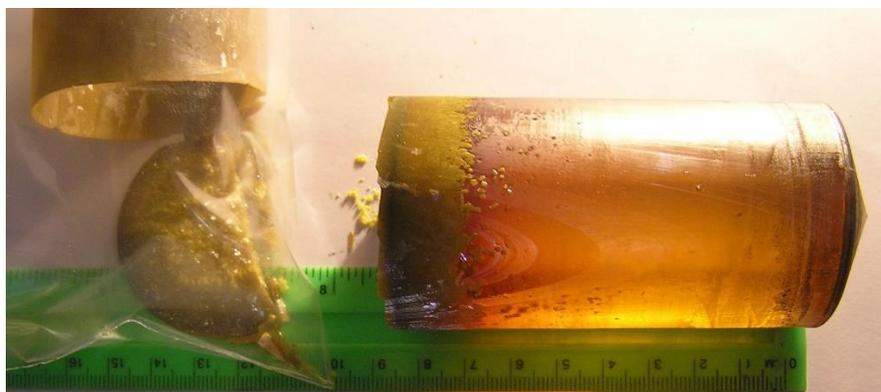

Fig. 3 Typical $Zn^{82}Se$ crystal ingot as grown by ISMA Kharkiv. On the left side (collected in a plastic bag) is the evaporated material deposited inside the crucible on the upper side of the walls.

### 4.3.1. Crystal radio-purity

The radio-purity of $Zn^{82}Se$ crystals is a parameter of crucial importance for 0vDBD application. Given the rareness of this hypothetical process, the experimental conditions have to ensure a background level below $1 \times 10^{-3}$ counts/keV/kg/y which asks for extremely low (ppt and less) acceptance limits of radioactive contamination elements in the crystal. Table 5 gives the concentration limits (ICP-MS measurement) for radioactive isotopes requested for the materials used in the production of $Zn^{82}Se$ crystals. The limits were fixed according to a background goal of $1 \times 10^{-3}$ counts/keV/kg/y for the LUCIFER project [8].

**Table 5** Concentration limits (ICP-MS certified) for radioactive isotopes requested for raw materials, reagents, consumables and intermediary products used for the production of $Zn^{82}Se$ crystals.

| Material | contamination limits [ppb] |
|---|---|
| elemental Zn and Se | $^{238}U < 0.2$ <br> $^{232}Th < 0.2$ |
| water and solvents for cleaning | $^{238}U < 2*10^{-3}$ <br> $^{232}Th < 2*10^{-3}$ |
| ZnSe powder synthesized and conditioned for crystal growth | $^{238}U < 0.2$ <br> $^{232}Th < 0.2$ |
| ZnSe crystal, ready-to-use | $^{238}U < 3*10^{-4}$ <br> $^{232}Th < 3*10^{-4}$ |
| $SiO_2$ powder for crystal polishing and textile polishing pads | $^{238}U < 4*10^{-3}$ <br> $^{232}Th < 4*10^{-3}$ |
| gloves, plastic bags, cleaning tissues, etc used for handling and package | $^{238}U < 4*10^{-3}$ <br> $^{232}Th < 4*10^{-3}$ |

### 4.3.2. Crystal growth and crystal perfection

The high purity of raw materials and careful conditioning of $Zn^{82}Se$ powder in view of crystal growth were the main reasons that have allowed us to achieve the production of $Zn^{82}Se$ crystals with a high crystal perfection in a relatively large scale production process. Special efforts were devoted to avoid the presence of very small (100μm) black-colored inclusions or larger (~1mm) dark clouds. The total absence of such inclusions and possible bubbles was guaranteed by the thermal regime (thermal gradients and lowering speed) applied during the whole growing process. Last but not least, the cleaning procedure and conditioning of the crucible has been fundamental to avoid the contamination of the melt by material condensed on the crucible walls. The modified geometry of the crucible mentioned above was also made for this purpose. Figure 4a shows the scanning electron microscope (SEM) images of a typical inclusion when present as very small isolated dust grain. The image at the right side of the figure is obtained with the BSD detector. The BSD characterization permits the differentiation between different elements (different atomic number) and allows to perform a qualitative evaluation of composition. The image proofs the homogeneous constitution of the grain. The precise composition of it measured by EDX analysis shows that the inclusion is composed mainly of carbon (61%) and oxygen (30%). Undoubtedly the oxygen was not present in the as-grown crystal. The presence of oxygen is due to the diffusion in the sample which was cut and polished, not clived from the as grown crystal. Other impurities detected inside the grain are zinc (3%), chlorine (2%), indium (2%), sulfur (1%) and other (<1%). The values measured outside the grain are: zinc (56.60%), selenium (41.54%), other (<1%). Values in parenthesis are normalized concentration in weight percent of the element. When present in the form of relatively large dark clouds (hundreds of micron typically), the inclusions proved to be in fact clusters of small inclusions mainly composed also of carbon. Figure 4b gives the SEM images together with the EDX element mapping image of a typical "dark cloud" inclusion.

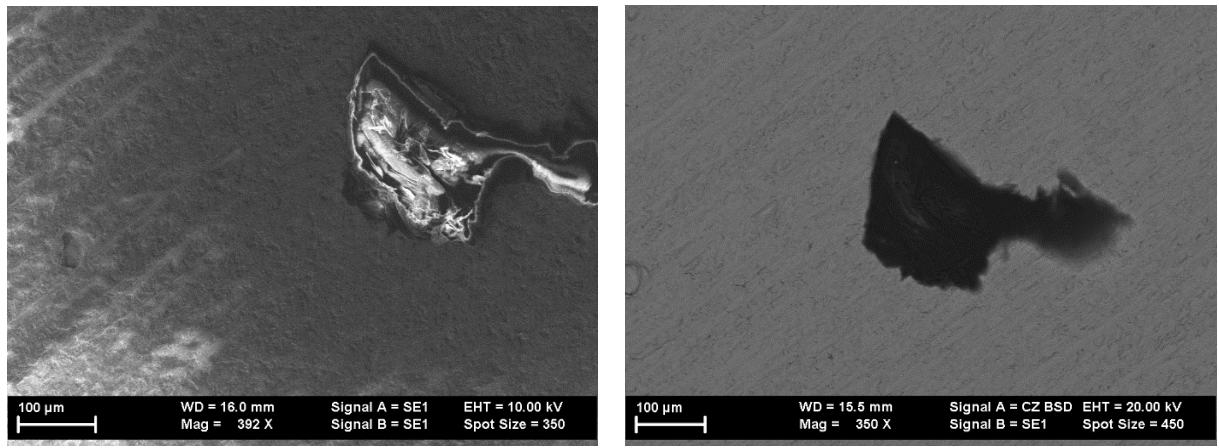

Fig.4a Scanning electron microscope (SEM) images of a typical "small" inclusion in ZnSe crystals. On the left, the "black point" visible also with naked eye proves to be a small (100-200μm) chop of impurity which was embedded in the crystal during the growth process. The image on the right is obtained with the BSD detector and proofs the homogeneous elemental composition of the inclusion.

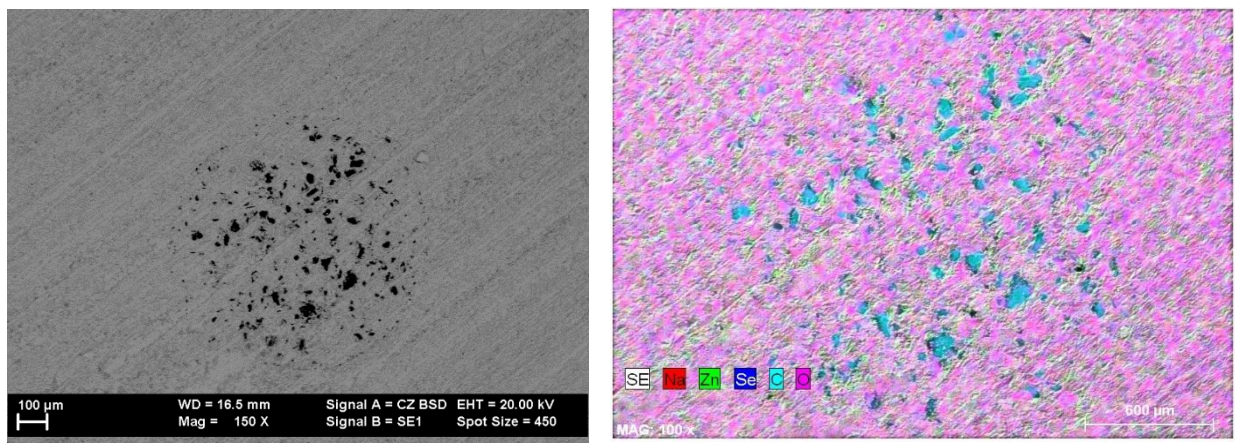

Fig.4b Scanning electron microscope (SEM) image (left) of a relatively large (~800μm) "dark cloud" inclusion. On the right, the EDX element mapping image showing that the dark cloud is in fact a cluster of small inclusions mainly composed of carbon.

The crystals grown in this work were subject to a special thermal regime aimed at reducing also the presence of veils and other extended inhomogeneities leading to cracks. Crystals homogeneity was guaranteed by the very high stability of the thermal regime during the crystal growth and the post-growth thermal regime. The thermal stability is also responsible for the absence of cracks. The start of the growth process was also tuned in order to avoid twinning and further development of a multiple-crystal ingot. Last but not least, the crystal cooling process was conducted under conditions that did not alter the crystal perfection during the phase transition from wurtzite to sphalerite symmetry. The fig. 5 shows the uniform orientation of crystalline plans of the as-grown $Zn^{82}Se$ crystals described in this work.

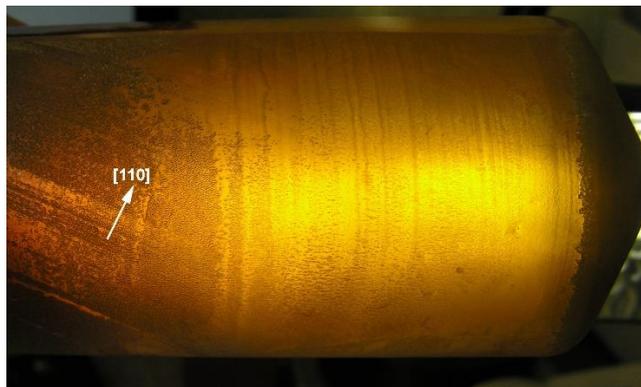

Fig. 5 As-grown $Zn^{82}Se$ crystal ingot showing uniform orientation of crystalline plans (arrow indicates the direction perpendicular to the 110 plane of the ZnSe crystal)

### 4.3.3. Optical properties of $Zn^{82}Se$ crystals

As already mentioned, ZnSe crystals have a very large transmission range, broader than silicon and germanium, extending from visible to the IR region of the spectrum. Nevertheless, optical absorption in ZnSe crystals can be strongly perturbed by impurities. Also, in the case of crystals grown from melt, due to the relatively high temperature, inhomogeneities may appear in the crystal bulk which are source of random wavering of the fundamental absorption edge. Figure 6 illustrates the transparency of an optically polished 50 mm long, 50 mm diameter $Zn^{82}Se$ crystal produced in this work. In figure 7a is given the transmission spectrum in the visible region for different ZnSe crystals grown in the phase of tuning of the production parameters. Characteristic for a crystal with a high concentration of defects is the crooked shape of the fundamental absorption edge at typically 550 nm. The presence of impurities also may harm the transmittance in the infrared region of the absorption spectrum, usually dominated by the absorption on free carriers [26]. Impurities of carbon compounds and selenium can be identified by the vibrational absorption bands in the infrared spectra [26]. Particularly, the iron impurities proved to worsen the bolometric performance of ZnSe crystals due to 3.3µm characteristic absorption band in the IR region of the absorption spectrum and consequent damping of phonon transport i.e. worsening of bolometric signal. The concentration of iron can be calculated from the absorption band at 3.3 µm (fig. 7b) and values of <1ppm were found to be compatible with a good bolometric performance. Figure 7 gives the optical transmission spectra of the preliminary series of ZnSe crystals with different types of defects and degrees of point defect concentration. The correlation between optical parameters and bolometric performance of ZnSe crystals is given in table 6.

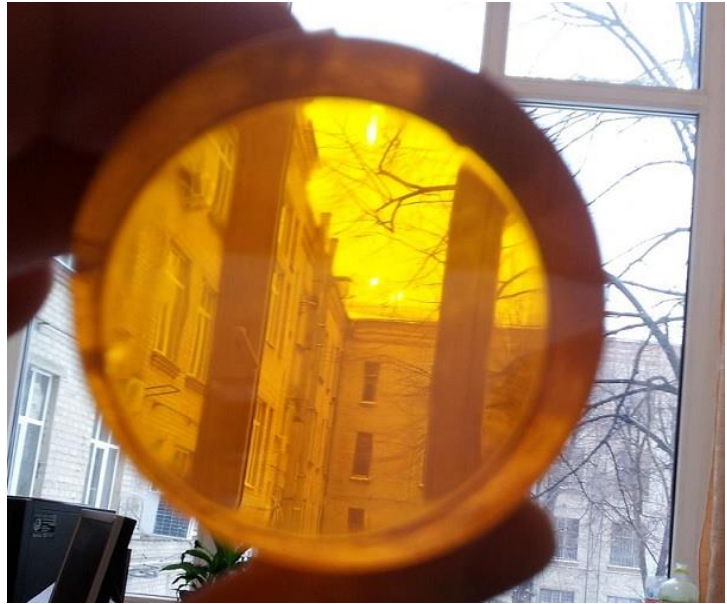

Fig. 6 Optically polished Zn$^{82}$Se crystal

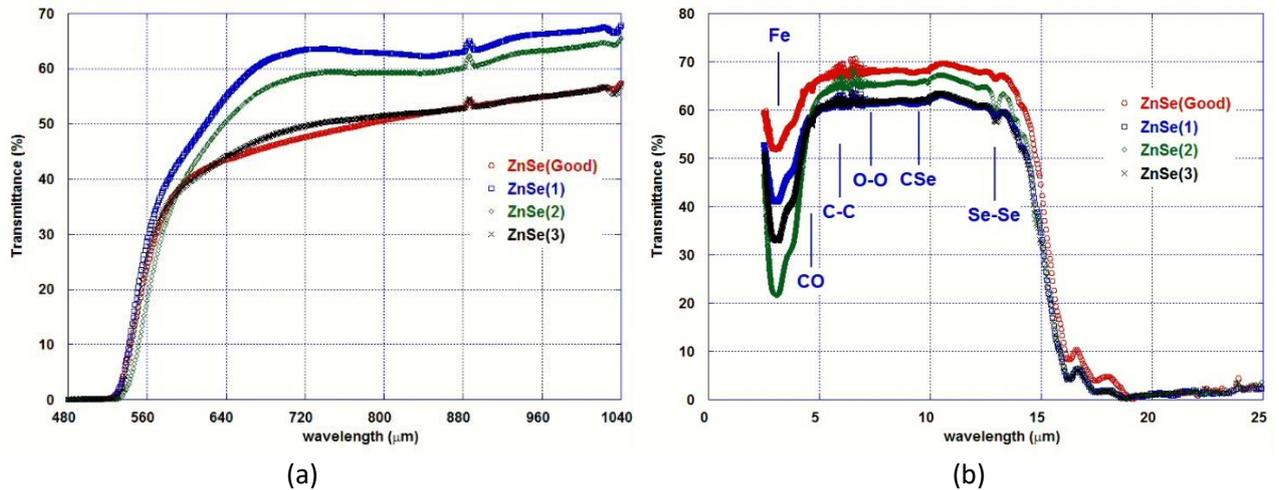

(a) (b)

Fig. 7. Visible (a) and infrared (b) spectra of ZnSe crystals Φ50 mm; L50 mm. The nature of defects responsible for the absorption bands in the infrared region of the spectrum are indicated with arrows.

Table 6 Bolometric performance and optical characteristics of preliminary series of ZnSe crystals

| sample ID | thermal pulse amplitude μV/MeV | point defects concentration | impurities concentration | LY (β/γ) keV/MeV) | Fe concentration [ppm] |
|---|---|---|---|---|---|
| 1 | 4 | high | moderate Se$_2$, CSe | 2.6 | 2 |
| 2 | 6 | high | high Se$_2$, CSe | 2.6 | 5 |
| 3 | 35 | moderate | high Se$_2$, CSe | 6.1 | 3 |
| G | 10-50 | moderate | moderate Se$_2$, CSe | 6.41 | 1.2 |

### 4.3.4. Cryogenic test

For a preliminary test of the crystals produced in this work, three enriched Zn$^{82}$Se crystals of the first production batch were operated as scintillating bolometers in the cryogenic facility of the hall-C of LNGS. The crystals, surrounded by 3M VM2002 reflecting foils, were assembled in a single array, interleaved by cryogenic light detectors (for a complete detector description see [27]). For technical reasons during the test, the cryostat was operated at a temperature of slightly above 20 mK, limiting the performances of the temperature sensor of the crystal (bolometric signal). At this temperature, indeed, the working resistance of sensors ranged from 0.17 to 0.22 MΩ, while the electronics is optimized for sensor resistances of tens of MΩ, corresponding to a temperature of about 5-10 mK. The electronic noise produced by the higher temperature dominated each detector baseline resolution that was characterized by different values according to the sensor resistance. Table 7 gives the detector baseline resolutions and light yield (LY) of the scintillating bolometers built using the preliminary data obtained for the three Zn$^{82}$Se crystals. In spite of the not most suitable operation conditions, the energy resolution in the region of interest for the expected 0vDBD of $^{82}$Se, was the same among the three crystals, proving the reproducibility of the bolometric performance. The performances of the crystals are expected to further improve with a lower base temperature of the cryogenic apparatus [28, 29].

**Table 7** Detector baseline resolutions and light yield (LY) of the scintillating bolometers made from three Zn$^{82}$Se crystals

| sample ID | Baseline Energy Resolution [keV FWHM] | Resolution [keV FWHM] | LY (β/γ) [keV/MeV] | LY (α) [keV/MeV] |
|---|---|---|---|---|
| Zn$^{82}$Se-1 | 7.0 | 30.1 ± 1.7 | 5.2 | 14.1 |
| Zn$^{82}$Se-2 | 14.1 | 29.7 ± 1.4 | 3.3 | 9.1 |
| Zn$^{82}$Se-3 | 18.6 | 30.2 ± 1.7 | 4.6 | 13.7 |

**Table 8** Activities in µBq/kg of the most critical isotopes measured for the three Zn$^{82}$Se crystals subject to the cryogenic test

|  | Zn$^{82}$Se-1 | Zn$^{82}$Se-2 | Zn$^{82}$Se-3 |
|---|---|---|---|
| $^{232}$Th | 13 ± 4 | 13 ± 4 | < 5 |
| $^{228}$Th | 32 ± 7 | 30 ± 6 | 22 ± 4 |
| $^{238}$U | 17 ± 4 | 20 ± 5 | <10 |
| $^{234}$U and $^{226}$Ra | 42 ± 7 | 30 ± 6 | 23 ± 5 |
| $^{230}$Th | 18 ± 5 | 19 ± 5 | 17 ± 4 |
| $^{210}$Pb | 100 ± 11 | 250 ± 17 | 100 ± 12 |

The scintillation light emitted by each Zn$^{82}$Se bolometer was measured using two cryogenic Ge light detectors [27], placed on the top and bottom surface of the cylindrical crystals. The light measured by the two light detectors after an energy deposit of 1 MeV ranges from 3.4 to 4.6 keV/MeV for electrons and from 9.2 to 14.4 keV/MeV for α's. Moreover, the different time development of light pulses produced by electrons and α interactions allows to completely disentangle and reject the α background [26].
The cryogenic test allowed to assess the radioactive contaminations of the crystals bulk with high sensitivity. The activities of the most dangerous isotopes are reported in µBq/kg in table 8. These contaminations are expected to produce a background in the energy region of interest of 1x10$^{-3}$ counts/keV/kg/y which is in compliance with the experimental requirements. Details on the results of the cryogenic tests are given in [27].

### 4.4. ZnSe crystal cutting, surface processing and handling

The as-grown Zn$^{82}$Se crystal ingots were cut and polished to the standard (LUCIFER) dimensions: 45mm diameter, 55mm length or less, depending on the quality of the ingot. The remaining material after ingot cut is to be recycled as indicated in fig. 1 (steps 15-17-14). The expected mass of a Zn$^{82}$Se crystal with standard dimensions is 469.976 g which corresponds to 248.162 g content of $^{82}$Se. The cutting was made using a precision diamond wire saw model STX-202A with three-dimensional digital control. The yield of the cutting process (taking into account the recovery of the scraps) was 96,72%.
The shaping and optical polishing of the crystals was made using commercial waterproof abrasive paper of different roughness. The shaping and optical polishing performed in normal laboratory conditions (above ground) were followed by an ultraclean surface processing consisting in two steps: a) chemical etching made in a clean room using 10% diluted HCl; b) polishing of the whole surface in a radon free clean room (underground) applying the polishing technique used for the TeO$_2$ crystals used in CUORE experiment and described in [10]. The final mechanical processing was carried out in order to guarantee a deep cleaning the crystal's surfaces, which may have been contaminated during the rough mechanical processing (cutting, shaping, grinding and lapping). The cleaning process is made in two steps, first by washing with ultrapure water and second by polishing. The targeted number of atomic layers to be taken away by this final polishing is on the order of $10^5$ to eliminate all impurity atoms that may have been adsorbed on the crystal's face and further diffused in its bulk. Finally, the mean depth of the crystal layer taken away by polishing was 10-20 μm and the least value of this width is 12μm which makes more than $10^5$ atomic layers. The relatively large spread of the polishing depth is due to the fact that polishing is not only a "surface cleaning" process.

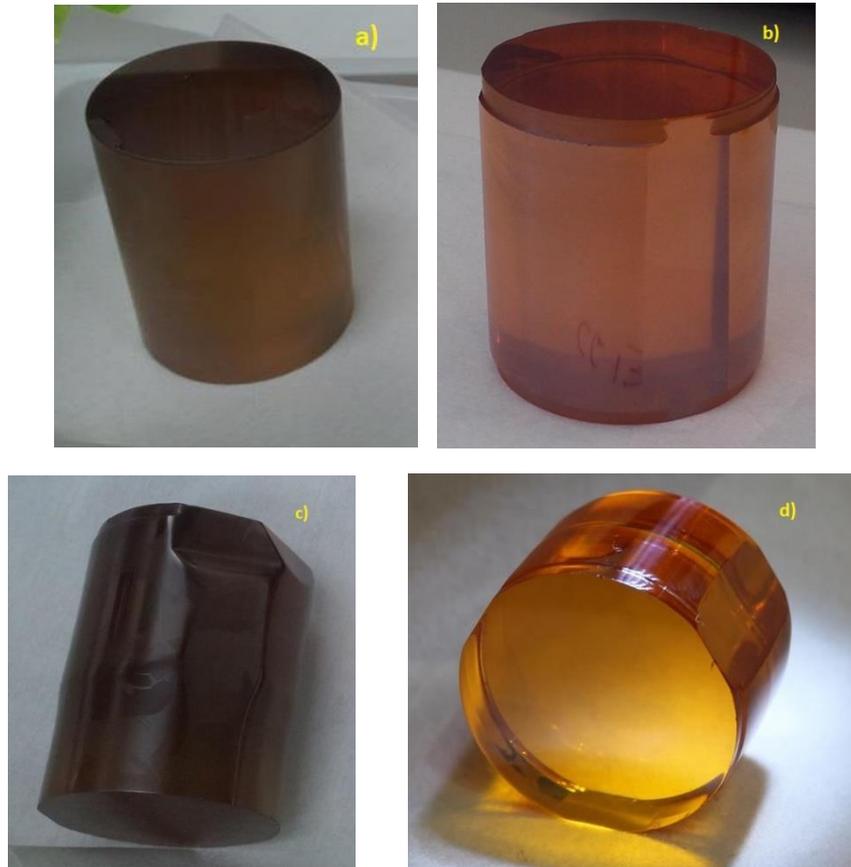

Fig. 8 Zn$^{82}$Se crystals of different shape ready for mounting: a) perfect cylindrical shape; b) collar and lateral strip aimed at fitting the copper frames; c) dedicated cut for maximum mass; d) short crystal

Besides removing a surface layer possibly contaminated with radioactive isotopes, the polishing is also intended to bring the crystal dimensions as close as possible to the nominal values and to improve the surface quality by removing possible extended defects induced by chemical etching. Special care was taken to avoid any contamination risk during the polishing operations performed inside the radon-suppressed clean-room belonging to the infrastructures of the DarkSide experiment at LNGS. The polishing suspension was prepared on site using $SiO_2$ powder (Admatechs product Admafine SO-E5) and ultrapure water added directly inside the plastic bags in which the $SiO_2$ was delivered. In this way, the crystals arrived in the clean room, came into direct contact only with materials previously certified for their radio-purity: ultra clean gloves, ultra-pure water, polishing suspension, polishing pads, cleaned and conditioned polyethylene sheets and vacuum plastic bags. The abrasive powder was selected for CUORE experiment based on dedicated research on the radio-purity and mechanical characteristics of different materials available on the market. The Admatechs product Admafine SO-E5 (average particle diameter 1.5μm) was selected for having the best radio-purity, even though it has moderate material removal efficiency. A special package made in a dedicated clean area and using certified polyethylene bags is guaranteed at the production site in Nagoya (Japan). The package consists of two successive bags. After removal of the outer bag in the clean room, the inner bag was used for the preparation of the polishing slurry by adding ultrapure water. At the end of clean room operation, crystals were packed in a triple vacuum package and stored in polyethylene vacuum boxes similar to those used for CUORE experiment [10]. The yield of the surface processing procedures (shaping and polishing) was 99,80%. An improvement of the global yield of the mechanical processing (cutting and polishing) is expected due to the foreseen recovery of $^{82}$Se from the waste which was carefully collected and stored at LNGS. Since not all crystals were grown in the same type of crucible and some of the as-grown ingots were not homogeneous a compromise was necessary in order to get the maximum possible of experimental mass. As a result, only a few crystals have a perfect cylindrical shape (fig. 8-a) while other have a special geometry aimed at fitting the copper frames (fig. 8-b) and some of them were subject to a dedicated cut in order to extract the largest possible mass from the as-grown ingot (fig. 8-c). Also in some cases very short crystals were extracted from the as grown ingot due to cracks occurred during the cutting process (fig. 8-d).

**Conclusion**
We have successfully grown high purity Zn$^{82}$Se crystals to be used as scintillating bolometers for the search of 0vDBD of $^{82}$Se. Dedicated methods and equipment were used along all steps of the production process in order to obtain crystals with excellent crystallographic characteristics, reflected by their very good energy resolution as bolometers (FWHM of the order of 30 keV in the spectral region of interest for 0vDBD of $^{82}$Se). The scintillation characteristics of crystals are also promising, the light yield of typically 5 keV/MeV for β/γ and 10 keV/MeV for α excitation will guarantee the identification of the nature of the interacting particle thus contributing to achieve a zero background level in the region of interest for double-beta decay investigation of $^{82}$Se. High sensitivity measurements of radio-isotope concentrations in raw materials, reactants, consumables, ancillaries and intermediary products and the application of very strict production and certification protocols led to a very low radio-contamination level measured in the ready-to-use Zn$^{82}$Se crystals. The cryogenic test performed in conditions similar to those foreseen for the 0vDBD experiment showed a radio-active contamination of the crystals of the order of few µBq/kg for the most critical isotopes which is compatible with the challenging requirements of 0vDBD experiments.

**Role of the funding sources**
This work was made in the frame of LUCIFER experiment funded by the European Research Council under the European Union Seventh Framework Programme (FP7/2007-2013) / ERC grant agreement n. 247115.

**Acknowledgements**
Authors are grateful to DarkSide collaboration for putting at our disposal their radon-suppressed clean-room at LNGS for the final polishing of our Zn$^{82}$Se crystals.